\documentclass[aps,superscriptaddress,twocolumn,twoside,floatfix,pra,nofootinbib,a4paper]{revtex4}
\usepackage{graphicx,amsmath,amssymb,color}
\usepackage[normalem]{ulem}
\usepackage{amsfonts}
\usepackage[ocgcolorlinks,colorlinks=true,linkcolor=blue,citecolor=red]{hyperref}
\usepackage{latexsym}
\usepackage{amsfonts}
\usepackage{mathrsfs}
\usepackage{natbib}
\usepackage{color,verbatim}
\DeclareMathAlphabet{\mathrsfs}{U}{rsfs}{m}{n}
\DeclareMathAlphabet{\mathpzc}{OT1}{pzc}{m}{it}
\DeclareMathAlphabet{\matheus}{U}{eus}{m}{n}
\DeclareMathAlphabet{\mathbbold}{U}{bbold}{m}{n}

\newtheorem{theorem}{Theorem}

\newcommand{\ba}{\begin{eqnarray}}
\newcommand{\be}{\begin{equation}}
\newcommand{\ee}{\end{equation}}

\newcommand{\ea}{\end{eqnarray}}
\newcommand{\ban}{\begin{eqnarray*}}
\newcommand{\ean}{\end{eqnarray*}}
\newcommand{\Tr}{\operatorname{tr}}

\begin{document}

\title{Quantum Correlations in Connected Multipartite Bell Experiments}

\author{Armin Tavakoli}
\affiliation{Department of Physics, Stockholm University, S-10691 Stockholm, Sweden}
\affiliation{ICFO-Institut de Ciencies Fotoniques, Mediterranean Technology Park, 08860 Castelldefels (Barcelona), Spain}
\affiliation{Computer Science Division, University of California, Berkeley, California 94720, USA}

%%%%%%%%%%%%%%%%%%%%%%%%%%%%%%%%%%%%%%%%%%%%%%%%%%%%%%%%%%%%%%%%%%%

\date{\today}
%Idea: April 2014 at ICFO Barcelona.

%%%%%%%%%%%%%%%%%%%%%%%%%%%%%%%%%%%%%%%%%%%%%%%%%%%%%%%%%%%%%%%%%%%

\begin{abstract}
Quantum correlations arising in Bell experiments, involving a physical source that emits a quantum state to a number of observers, have been intensively studied over the last decades. Much less is known about the nature of quantum correlations arising in network structures beyond the Bell experiments. Such networks can involve many independent sources emitting states to observers in accordance with the network configuration. Here, we will study classical and quantum correlations in a family of networks which can be regarded as compositions of several independent multipartite Bell experiments connected together through a central node. For such networks we present tight Bell-type inequalities which are satisfied by all classical correlations. We study properties of the violations of our inequalities by probability distributions arising in quantum theory. 
\end{abstract}

%%%%%%%%%%%%%%%%%%%%%%%%%%%%%%%%%%%%%%%%%%%%%%%%%%%%%%%%%%%%%%%%%%%

%Quantum cryptography

\maketitle

\section{Introduction}
\begin{figure}
\centering
\includegraphics[width=\columnwidth]{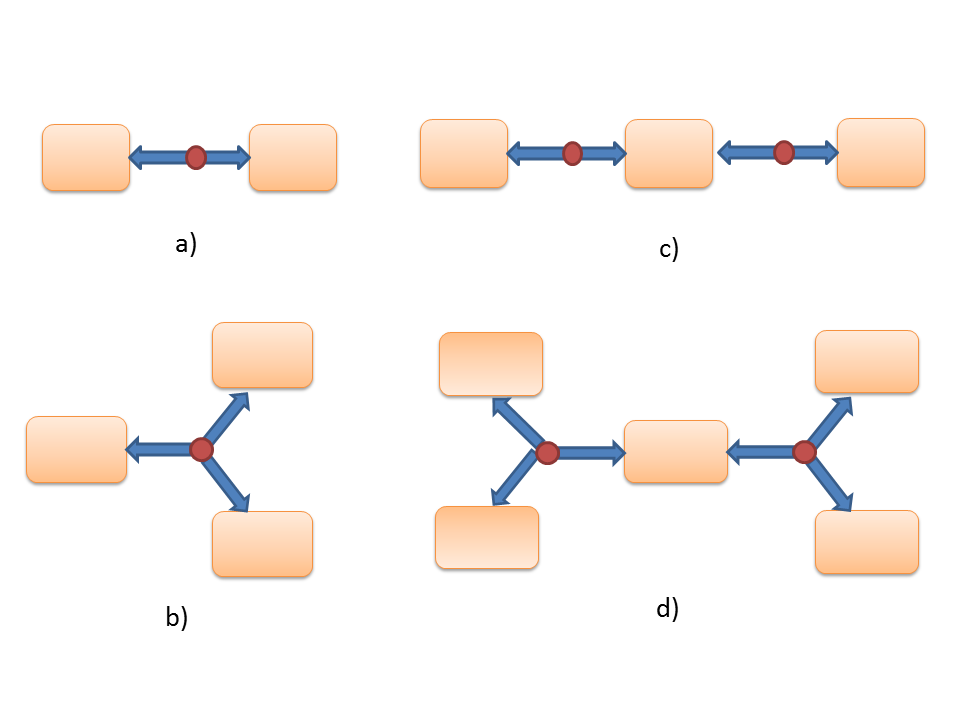}
\caption{Different network configurations: a) bipartite Bell experiment , b) multipartite Bell experiment, c) chain network of three observers, d) two connected three-partite Bell experiments ($\mathscr{N}_2^2$).}
\label{BellExp}
\end{figure}

Statistical correlations between outcomes obtained in different measurement events can provide insight to the physical causes of the statistics. One example is the celebrated theorem of John Bell which shows that statistical correlations arising in quantum theory cannot be explained by any theory that respects the principles of locality and realism \cite{Bell64}. The bipartite Bell experiment, see figure \ref{BellExp} a), considers two observers Alice and Bob each performing measurements $x$ and $y$ respectively, that are randomly chosen from some set of possible measurement setting, in space-like separated measurement events on a shared state. The measurements return outcomes $a$ and $b$ for Alice and Bob respectively. If the resulting probability distribution $P(a,b|x,y)$ respects the principle of locality, then no influence can propagate fast enough between the measurement events in order for the outcomes to directly influence each other i.e. all correlations between the outcomes of Alice and Bob must be due to some cause $\lambda$ originating from the common past of the two particles. Furthermore, if $P(a,b|x,y)$ also respects realism, the physical properties of the system are well-defined before a measurement takes place. Correlations satisfying such a description are deemed classical and can be written on the form    
\begin{equation}\label{LC}
P(a,b|x,y)=\int d\lambda q(\lambda) P(a|x,\lambda)P(b|y,\lambda)
\end{equation}
where $q$ is some probability density function. However, if $P(\cdot)$\footnote{The notation $P(\cdot)$ will be used as an abbreviation for a probability distribution when the argument is clear from context. E.g. in this case $P(\cdot)=P(a,b|x,y)$} is given by quantum theory then it may not admit the above form and is thus deemed intrinsically quantum. The properties of quantum correlations in such bipartite Bell experiments have been thoroughly studied \cite{Bell14}. 

A straightforward extension of the bipartite Bell experiment is the multipartite Bell experiment in which a source emits a many-particle state and each particle is measured at a different measurement event, see figure \ref{BellExp} b). An interesting property of the arising multipartite quantum correlations is that they can exponentially outperform the classical bound on Bell inequalities, as was first demonstrated for  Mermin's inequality \cite{M90}.
 
However, the properties of quantum and classical correlations in more sophisticated network configurations than the standard Bell experiments are to great extent unknown. Such networks can involve multiple independent sources each distributing a state to some set of observers performing randomly chosen measurements. Despite the initial independence of the involved sources, suitably chosen measurements can give rise to strong quantum correlations spanning the whole network.

Quantifying correlations in network structures was first considered for the network consisting of three observers in a chain configuration with two independent sources \cite{BGP10, BRGP12}, see figure \ref{BellExp} c). Bell-type inequalities, and their quantum violations, for such chain networks involving arbitrary many observers have been studied in Ref.\cite{MPS15}. Another class of networks with star-shaped configuration i.e. configurations involving many bipartite sources connected by a center node, were studied in  Ref.\cite{TSCA14}. Recently, it was shown that given knowledge of a Bell-type inequality for a network, one can recursively derive another Bell-type inequality for the same network but with one added source connected to one observer \cite{RB15}. Advances on a method for deriving Bell-type inequalities for networks has been made in \cite{C15}. Furthermore, correlations in network structures have been more broadly studied through the lens of causal inference \cite{CAP10, F12(1), F12(2), LS13, CLM14, HLP14, CK15, LS15, WS15, RMG15}. 

There are several motivations for studying quantum correlations in networks beyond the Bell experiments. Firstly, from a fundamental point of view it is interesting to understand the nonclassical  properties of quantum correlations. The exploration of physical correlations on networks is in principle very different from the analog case in Bell experiments. The reason for this is that the notion of classical correlations in a network naturally leads to constraints stronger than the assumption \eqref{LC} associated to classical correlations in Bell experiments. The intuition behind this is that in Bell experiments the observes always share a common local random variable $\lambda$ which allows any two observers to be directly correlated with each other. This is no longer the case on networks i.e. randomness is not shared between all pairs of observers. Therefore, correlations on networks are typically subjected to stronger constraints than correlations in Bell experiments. To what extent these constraints limit the strength of classical and quantum correlations, and the relation between the two, is of evident fundamental interest. Secondly, it is well-established that quantum correlations have many applications in quantum information e.g. in cryptography \cite{DI} and randomness generation \cite{Rand}. On the one hand, most implementations of such protocols involve a small number of observers connected by a single source. On the other hand, large-scale multiuser quantum communication networks are arguably one of the main objectives of applied quantum information. It is therefore relevant to study the ability of networks to support quantum correlations. Examples in which network structures are important inclulde entanglement swapping experiments \cite{ZZHE93}, entanglement percolation \cite{ACL07} and quantum repeaters protocols \cite{SSRG11, SB05} aiming to tackle the experimental challenges associated with distributing quantum states over large distances. Thirdly, it is known that the ability of a quantum state to violate a Bell inequality can be activated by considering many copies of the same state distributed in a network structure \cite{CASA10}. Thus, quantum correlations on networks can yield advantages over those in Bell experiments. However, this result was shown without invoking the stronger constraints associated to networks i.e. the independence of the sources. It is of evident interest to search for such advantages when imposing this stronger constraint. 
  
In this work, we will explore classical and quantum correlations on a class of networks consisting of many, initially independent, multipartite Bell experiments that are all connected through a central node, see figure \ref{BellExp} d). These networks can also be viewed as generalizations of the networks considered in \cite{BGP10, BRGP12, TSCA14} to scenarios involving many multipartite sources. We will derive Bell-type inequalities for such networks and study the properties of their quatum violations. In the light of the above motivations for studying quantum correlations on networks, we are in particular interested in searching for advantages over standard Bell experiments.

\section{Classical Correlations in Connected Bell Experiments}

The networks we will investigate are composed of $n$ sources, each distributing a state of $L+1$ particles in such a way that for each source $L$ distinct observers receive one particle each and the final particle is sent to a particular observer (called Bob) that acts as the center node connecting the $n$ Bell experiments. Thus, Bob will have $n$ particles at his disposal while the other $n\times L$ observers each hold one particle. We can index the described network by the pair $(n,L)$ and we abbreviate the network configuration by $\mathscr{N}_n^L$. For example, figure \ref{BellExp} d) represents the network $\mathscr{N}_2^2$ since there are two independent sources and each distribute a $2+1$ particle state.

Each observer (except Bob) will randomly choose one of two local measurements. The measurement choice of the $k$'th observer associated to the $j$'th source is denoted $x_j^k\in\{0,1\}$ and the associated outcome is denoted $a_j^k\in\{0,1\}$. Bob will randomly select a measurement $y\in\{0,...,2^L-1\}$ from which he will output $b\in\{0,1\}$. One can also consider variations in which Bob performs a measurement which returns more than two possible outcomes.

\subsection{Defining classical correlations}

Let us begin with defining the notion of classical correlations in $\mathscr{N}_n^L$. The natural extension of the classical probability distribution in the standard bipartite Bell experiment \eqref{LC}  is as follows: a probability distribution in $\mathscr{N}^{L}_n$ is classical if it is a mixture of local outcomes that depend only on the local measurement setting and the relevant physical causes rendering the associated outcome deterministic i.e.,
\begin{equation}\label{LocalRealism}
P(\overline{a},b|\overline{x},y)\!=\!\int \!\prod_{j=1}^{n}\left(\!d\lambda_j q_j(\lambda_j)\prod_{k=1}^{L}\left(P(a_j^k|x_j^k,\lambda_j)\right)\!\right)\!\! P(b|\overline{\lambda},y)
\end{equation}  
where we have by $q_j$ denoted the probability distribution function associated to the physical cause $\lambda_j$ associated to the $j$'th source. Also, we will frequently make use of the bar-notation to denote a collection of associated variables e.g. $\overline{a}=\left(a_1^1,\ldots,a_1^{L},\ldots, a_n^{L}\right)$ and similarly for $\overline{x}$ and $\overline{\lambda}$.

\subsection{Bell-type inequalities for $\mathscr{N}_n^L$}

We will now derive a family of Bell-type inequalities for the network $\mathscr{N}_n^L$.

Introduce a set of quantities $\{K_X\}_X$ indexed by $X$  which are linear combinations of conditional probabilities $P(\overline{a},b|\overline{x},y_X)$ arising in $\mathscr{N}_n^L$. The index $X$ runs over the  power set  (the set of all subsets), $\mathbb{P}$, of the set $\mathbb{N}_L=\{1,...,L\}$, i.e. every subset of $\mathbb{N}_L$ corresponds to a value of $X$ to which we associate a quantity $K_X$. For every element of $\mathbb{P}(\mathbb{N}_L)$, we define 
\begin{multline}\label{KX}
K_{X}=\frac{1}{2^{nL}}\sum_{\overline{x}}g(X)\sum_{\overline{a},b}(-1)^{b+\sum_{j,k}a_j^{k}} P(\overline{a},b|\overline{x},y_X),
\end{multline}
where in the expression  $P(\overline{a},b|\overline{x},y_X)$ we use the index $X$ in $y_X$ to identify the particular measurement of Bob (chosen from the $2^L$ possible settings) associated to the quantity $K_X$, and where the function $g(X)$ is defined as
\begin{eqnarray}
g(X)=\prod_{j=1}^{n} (-1)^{\sum_{k\in X}x_j^k}.
\end{eqnarray}

We will now state and prove the following theorem: 

\begin{theorem}
If given a probability distribution $P$ in $\mathscr{N}_n^{L}$ that admits a classical model, then the following inequality holds,
\begin{equation}\label{Inequality1}
\sum_{X \in \mathbb{P}(\mathbb{N}_{L})} |K_X|^{1/n} \leq 1.
\end{equation}
\end{theorem}
\textit{Proof:} Implementing the classical model \eqref{LocalRealism} in $\mathscr{N}_n^{L}$ with the quantities $K_X$ yields
\begin{multline}
K_X=\frac{1}{2^{nL}}\sum_{\overline{x}}g(X)\int \prod_{j=1}^{n}\Bigg[d\lambda_j q_j(\lambda_j)\\
\times\prod_{k=1}^{L}\left(\sum_{a_j^k}(-1)^{a_j^k}P(a_j^k|x_j^k,\lambda_j)\right)\Bigg]\sum_{b} (-1)^bP(b|\overline{\lambda},y_X).
\end{multline}
Introducing the following notations
\begin{eqnarray}
\langle B_{y_X}\rangle_{\overline{\lambda}}=\sum_{b} (-1)^bP(b|\overline{\lambda},y_X)\\
\langle A_{x^k_j}^{j,k}\rangle_{\lambda_j}=\sum_{a_j^k}(-1)^{a_j^k}P(a_j^k|x_j^k,\lambda_j)
\end{eqnarray}
and using the fact that $\left|\langle B_{y_X}\rangle_{\overline{\lambda}}\right|\leq 1$, the quantity $|K_X|$ can be bounded from above by
\begin{equation}\label{bound}
|K_X|\leq \prod_{j=1}^{n}\frac{1}{2^L}\int d\lambda_jq_j(\lambda_j)\prod_{k=1}^{L}
\left|\sum_{x_j^k}(-1)^{\delta^k_X x_j^k}\langle A_{x_j^k}^{j,k}\rangle_{\lambda_j}\right|,
\end{equation} 
where we have introduced the binary function $\delta^k_X=1$ if $k\in X$ and $\delta^k_X=0$ otherwise.

At this point, we need to introduce the following lemma: let $c_s^l$ be non-negative real numbers and $m,n$ be positive integers, then it holds that
\begin{equation}\label{lemma}
\sum_{l=1}^{m}\left(\prod_{s=1}^{n}c_s^l\right)^{1/n}\leq \prod_{s=1}^{n}\left(\sum_{l=1}^{m}c_s^l\right)^{1/n}.
\end{equation}
This was proven in Ref.\cite{TSCA14}: for every $l$ use that the arithmetic mean of the sequence $\{c_s^l\}_{s=1}^n$ is always larger than or equal to the geometric mean. Establishing such a relation for every $l$ and then summing the right- and left-hand sides over $l$ will prove the lemma. 

Applying the relation \eqref{lemma} to the inequalities \eqref{bound} with $c_s^l$ corresponding to each factor of the product series over $j$ in \eqref{bound}, we can construct the following inequality
\begin{multline}\label{ExpandedInequality}
\sum_{X \in \mathbb{P}(\mathbb{N}_{L})} |K_X|^{1/n}\leq \prod_{j=1}^{n}\Bigg[\frac{1}{2^L}\int d\lambda_j q_j(\lambda_j)\\
 \times \sum_{X\in \mathbb{P}(\mathbb{N}_{L})}\prod_{k=1}^{L} \left|\sum_{x_j^k}(-1)^{\delta^k_Xx_j^k}\langle A_{x_j^k}^{j,k}\rangle_{\lambda_j}\right|\Bigg]^{1/n}.
\end{multline} 
To find an upper bound on the sum over the product series in the integrand, we note that for a given $j$ and $X$ we can write the absolute value in the integrand as $r^{j,k}_{\pm}\equiv|\langle A^{j,k}_{0}\rangle_{\lambda_j}+(-1)^{\delta^k_X }\langle A^{j,k}_{1}\rangle_{\lambda_j}|$ where the $\pm$ index referrs to the sign inside the modulus determined by $(-1)^{\delta_X^k}$. The integrand consists of products of $L$ such factors over which a sum is taken so that all possible arrangements of the sign inside the absolute value of each factor in the product occurs. Therefore, we can factor the integrand into a product of $L$ factors as follows: $(r_{+}^{j,1}+r_{-}^{j,1})(r_{+}^{j,2}+r_{-}^{j,2})\ldots (r_{+}^{j,L}+r_{-}^{j,L})$. Since all $\langle A_{x^k_j}^{j,k}\rangle_{\lambda_j}$ are real and their modulus is bounded by $1$, it must be that $r_{+}^{j,k}+r_{-}^{j,k}\leq 2$ for any given $j,k$. Therefore, we conclude that for all $j$:  
\begin{equation}
\sum_{X\in \mathbb{P}(\mathbb{N}_{L})}\prod_{k=1}^{L} \left|\sum_{x_j^k}(-1)^{\delta_X^kx_j^k}\langle A_{x_j^k}^{j,k}\rangle_{\lambda_j}\right|\leq 2^{L}.
\end{equation}  
Implementing this upper bound with \eqref{ExpandedInequality}, and using that $\int q_j(\lambda_j)d\lambda_j=1$ for any $j$, we obtain that 
\begin{equation}
\sum_{X \in \mathbb{P}(\mathbb{N}_{L})} |K_X|^{1/n} \leq 1.
\end{equation}
\begin{flushright}
$\blacksquare$
\end{flushright}

Notice that the inequality \eqref{Inequality1} admits various special cases that have been derived in earlier work: when $(n,L)=(2,1)$ the inequality reduces to that found in Ref.\cite{BRGP12}, and the Bell-type inequality for the star-network found in Ref.\cite{TSCA14} is recovered by considering only bipartite sources ($L=1$).

\subsection{Tightness of the inequality} 
We will now state and prove the following theorem:
\begin{theorem}
The inequality \eqref{Inequality1} is tight in the sense that whenever a conditional probability distribution $P(\cdot)$ on $\mathscr{N}_n^{L}$ satisfies \eqref{Inequality1}, then $P(\cdot)$ necessarily admits a classical model. 
\end{theorem}
\textit{Proof:} We show tightness of \eqref{Inequality1} by explicitly constructing a classical model that continuously saturates the classical bound i.e. we find a family of classical strategies parametrized by a set of continuous variables that saturate the classical bound for given $(n,L)$.

Let $\overline{\alpha}$ be a string of $n\times L$ random bits i.e. for $j=1,...,n$ and $k=1,...,L$, there is a $\alpha_j^k\in\{0,1\}$, subject to the probability distribution $P(\alpha_j^k=0)=p_j^k$. Introduce a family of classical strategies $D(\{p_j^k\})$ which depend on the probability distribution of all the bits in $\overline{\alpha}$, 
\begin{equation}
D: \hspace{5 mm} a_j^k=\lambda_j\oplus \alpha_{j}^k x_{j}^k \hspace{7 mm} b=\left\{
     \begin{array}{lr}
       \bigoplus_{j} \lambda_j & \text{if L is odd}  \\
       0 & \text{if L is even}\\
     \end{array}
     \right.
\end{equation}
with the distribution of $\lambda_j$ being $q(\lambda_j=0)=\frac{1}{2}$.

For any $X\in \mathbb{P}(\mathbb{N}_{L})$ the strategy $D(\{p_j^k\})$ yields 
\begin{equation}
K_X=\prod_{j=1}^n\left(\prod_{k\in X} (1-p_j^k)\prod_{k\notin X}p_j^k\right).
\end{equation}
Let us denote the expression inside the bracket by $c_j^X$. We can then bound the left-hand-side of \eqref{Inequality1} by
\begin{multline}
\sum_{X \in \mathbb{P}(\mathbb{N}_{L})}|K_X|^{1/n}=\sum_{X \in \mathbb{P}(\mathbb{N}_{L})} \left(\prod_{j=1}^{n} c_j^X\right)^{1/n}\\
\leq \prod_{j=1}^{n}\left(\sum_{X \in \mathbb{P}(\mathbb{N}_{L})}c_j^X \right)^{1/n}=1
\end{multline}
where in the second step we have used the lemma \eqref{lemma} and in the last step used that $\sum_{X} c_j^X=\prod_{k=1}^{L}(p_j^k+(1-p_j^k))^{1/n}=1$ for every $j$. Importantly, the inequality in the second step becomes an equality if and only if we impose that $c_0^X=c_1^X=\ldots=c_n^X$ corresponding to constraining the probability distribution by $p^k\equiv p_0^k=p_1^k=\ldots=p_n^k$. Thus, the startegy $D(p^1,\ldots,p^L)$ achieves the bound of \eqref{Inequality1} for every given value of $\{p^k\}$. We did only consider the case in which all $K_X$'s are positive, but it is clear from the above symmetries that the tightness of the inequality is implied.

\begin{flushright}
$\blacksquare$
\end{flushright} 

\begin{figure}
\centering
\includegraphics[width=\columnwidth]{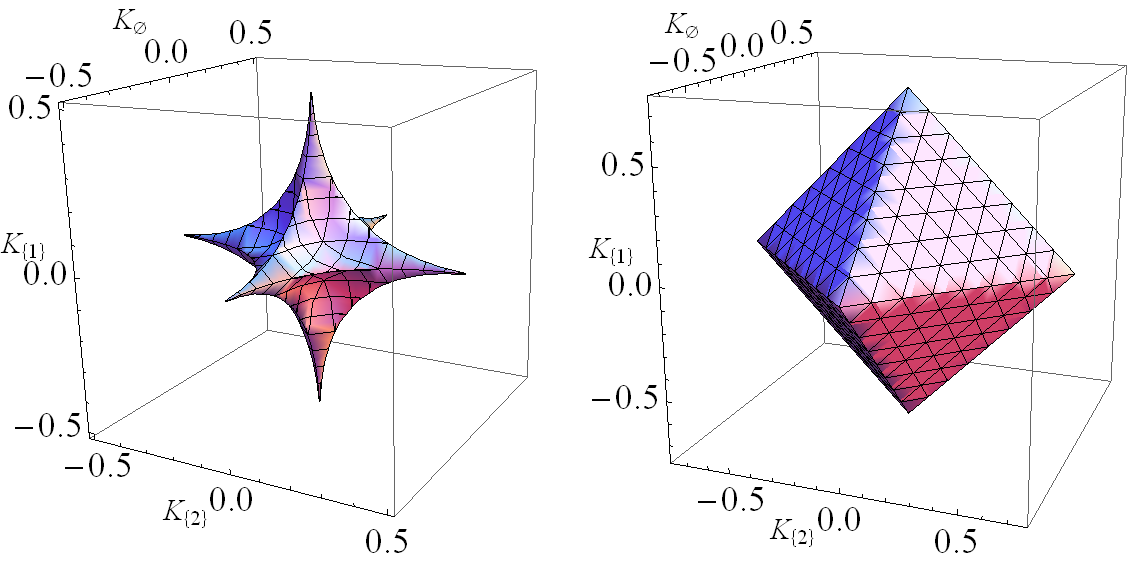}
\caption{Left: Non-convex subset of the classical set of correlations for $(n,L)=(2,2)$ obtained from fixing $K_{\{1,2\}}=1/16$. Right: Convex subset of the classical set of correlations for $(n,L)=(1,2)$ obtained from fixing $K_{\{1,2\}}=1/4$ .}
\label{nonconvex}
\end{figure}

Since we know that the inequality \eqref{Inequality1} is tight we can identify an often re-occuring important property for classical correlations in network structures: the set of points in the space $(K_\emptyset,...,K_{\{1,...,L\}})$ which satisfies the inequality \eqref{Inequality1}  is non-convex. We illustrate the non-convexity in figure \ref{nonconvex} (left sub-figure) in which we have set $(n,L)=(2,2)$ and plotted the subset of the region that satisfies \eqref{Inequality1} in $(K_{\emptyset},K_{\{1\}},K_{\{2\}},K_{\{1,2\}})$-space associated to $K_{\{1,2\}}=1/16$. To illustrate that the non-convexity arises from the network structure with multiple sources, we have also plotted a subset of the classical set arising in the standard multipartite Bell experiment corresponding to $(n,L)=(1,2)$ which is illustrated in figure \ref{nonconvex} (right sub-figure) for $K_{\{1,2\}}=1/4$.

\section{Quantum Correlations in Connected Bell Experiments}\label{sec3}
Quantum theory may predict violations of the inequality \eqref{Inequality1}. A quantum model for the probability distribution $P(\bar{a},b|\bar{x},y)$ takes the form 
\begin{multline}\label{pdist}
P(a_1^1\ldots a_n^{L} b|x_1^1\ldots x_n^{L}y_X)=\\
\Tr\left(\left(M_{a_1^1|x_1^1}\otimes\ldots \otimes M_{a_n^{L}|x_n^{L}}\otimes M_{b|y_X}\right) \left(\rho_{1}\otimes\ldots \otimes \rho_{n}\right)\right)
\end{multline}
where $M_{a_j^k|x_j^k}$ is the measurement operator associated to outcome $a_j^k$ when the measurement choice is $x_j^k$, and $\rho_{j}$ is the state of $L+1$ qubits emitted by the $j$'th source. The tensors are computed over the relevant Hilbert spaces (and the subspaces should therefore be rearranged accordingly).

Let Bob perform the measurements
\begin{eqnarray}\label{BobM}
M_{b|y}=\sum_{b_1\oplus\ldots \oplus b_n=b} \Pi_{b_1}^y\otimes \dots \otimes \Pi_{b_n}^y
%M_{b|1}=\sum_{b_1\oplus\ldots \oplus b_n=b} \Pi_{b_1}^1\otimes \dots \otimes \Pi_{b_n}^1,
\end{eqnarray}
for $y\in\{0,1\}$ where we define $\Pi_{b_i}^0=\frac{1}{2}\left(1+(-1)^{b_i}X\right)$ and $\Pi_{b_i}^1=\frac{1}{2}\left(1+(-1)^{b_i}Y\right)$ corresponding to the projectors onto the positive and negative subspaces of the Pauli operators 
\begin{equation}
X=\begin{pmatrix}
0 & 1\\
1 & 0
\end{pmatrix}
\hspace{15 mm}
Y=\begin{pmatrix}
0 & -i\\
i & 0
\end{pmatrix}.
\end{equation}
The action of Bob at the center node can be regarded as performing $n$ measurements, either all $X$ or all $Y$, with outcomes $b_1,\ldots,b_n$ and then process these $n$ outcomes into $b=b_1\oplus\ldots\oplus b_n$ which is announced as the final output. Hence, Bob's measurement is separable and the probability distribution \eqref{pdist} can be written
\begin{multline}\label{Pdist}
P(a_1^1\ldots a_n^{L} b|x_1^1\ldots x_n^{L}y_X)=\\
\sum_{b_1\oplus\ldots\oplus b_n=b} \prod_{j=1}^{n}\Tr\left(\left(\bigotimes_{k=1}^{L} M_{a_j^k|x_j^k}\otimes \Pi^{y_X}_{b_j}\right) \rho_{j}\right)\\
=\sum_{b_1\oplus\ldots\oplus b_n=b} \prod_{j=1}^{n} P(a_j^1\ldots a_j^{L}b_j|x_j^1\ldots x_j^{L}y_X)\\
=\sum_{b_1\ldots b_{n-1}}P(a_1^1\ldots a_1^{L}b_1|x_1^1\ldots x_1^{L} y_X)\times \dots \times \\
P(a_{n-1}^1\ldots a_{n-1}^{L}b_{n-1}|x_{n-1}^1\ldots x_{n-1}^{L} y_X)\times\\
 P(a_n^1\ldots a_n^{L}b\oplus b_1\oplus\ldots\oplus b_{n-1}|x_n^1\ldots x_n^{L} y_X).
\end{multline}
In the first equality we have used the linearity of the trace operation and that $\Tr\left(O_1\otimes O_2\right)=\Tr\left(O_1\right)\Tr\left(O_2\right)$. In the final equality we have expanded the product and rewritten the domain of the sum to go over $b_1,...,b_{n-1}$ by setting $b_n=b\oplus b_1\oplus...\oplus b_{n-1}$.

Let all the $n$ sources distribute the $(L+1)$-qubit Greenberger-Horne-Zeilinger (GHZ) state defined as $|GHZ\rangle\!\!=\!\!\frac{1}{\sqrt{2}}\left(|0\rangle^{\otimes L+1}+|1\rangle^{\otimes L+1}\right)$ i.e. $\rho_j\!=\!|GHZ\rangle\langle GHZ|$ for every $j$. If we let the measurements associated to observers acting on a single qubit be either $X$ (labelled $0$) or $Y$ (labelled $1$) the probability distribution obtained in every separate multipartite Bell experiment takes the form
\begin{multline}\label{P}
P(a_j^1\ldots a_j^{L}b_j|x_j^1\ldots x_j^{L} y_X)=\\
\frac{1}{2^{L+1}}\left[1+(-1)^{b_j+\sum_{k=1}^{L} a_j^k}\cos\left(\frac{\pi}{2}\left(\sum_{k=1}^{L} x_j^k+y_X\right)\right)\right].
\end{multline} 
Substituting this into \eqref{Pdist} we obtain the joint probability distribution
\begin{multline}
P(\overline{a},b|\overline{x},y_X)=
%\frac{1}{2^{\sum_{j=1}^{n}L_j}}\sum_{b_1\ldots b_{n-1}}\left(1+(-1)^{b_1+\sum_{k=1}^{L_1}a_1^k}\cos\left(\frac{\pi}{2}\left(\sum_{k=1}^{L_1}x_1^k+y\right)\right)\right)\times \dots \\
 %\times %\left(1+(-1)^{b_{n-1}+\sum_{k=1}^{L_{n-1}}a_{n-1}^k}\cos\left(\frac{\pi}{2}\left(\sum_{k=1}^{L_{n-1}}x_{n-1}^k+y\right)\right)\right)\times\\
 %\left(1+(-1)^{b+b_1+\dots+b_{n-1}+\sum_{k=1}^{L_{n}}a_n^k}\cos\left(\frac{\pi}{2}\left(\sum_{k=1}^{L_{n}}x_{n}^k+y\right)\right)\right)=\\
 \frac{1}{2^{n(L+1)}}\sum_{b_1\ldots b_{n-1}}\Bigg[1+\ldots+(-1)^{b+\sum_{j,k}a_j^k}\\
 \times 
 \prod_{j=1}^{n}\cos\left(\frac{\pi}{2}\left(\sum_{k=1}^{L}x_j^k+y_X\right)\right)\Bigg]	.
\end{multline}
The sum is taken over $2^{n}$ terms. We observe that all the terms not explicitly written out contain a factor on the form $(-1)^{b_i}$ and that this will cause all terms not explicitly written to be cancelled when executing the sum. In conclusion we are left with
\begin{multline}\label{prob}
P(\overline{a},b|\overline{x},y_X)=\\
\frac{1}{2^{nL+1}}\left(1+(-1)^{b+\sum_{j,k}a_j^k}\prod_{j=1}^{n}\cos\left(\frac{\pi}{2}\left(\sum_{k=1}^{L}x_j^k+y_X\right)\right)\right).
\end{multline}

Let us now compute the value of the left-hand-side of \eqref{Inequality1} under the above calculated probability distribution. By direct insertion of \eqref{prob} into the definition  \eqref{KX} of $K_X$, we find that
\begin{equation}\label{KXP}
K_{X} %\frac{1}{2^{nL}}\sum_{\bar{x}}g(X) \prod_{j=1}^{n}\cos\left(\frac{\pi}{2}\left(\sum_{k=1}^{L}x_j^k+y_{X}\right)\right)\\
=\frac{1}{2^{nL}}\sum_{\bar{x}} \prod_{j=1}^{n}(-1)^{\sum_{k' \in X}x_j^{k'}}\!\!\cos\left(\frac{\pi}{2}\left(\sum_{k=1}^{L}x_j^k+y_{X}\right)\right).\\
\end{equation}

First, we need to define how we assign values to $y_X$. We define the convention $y_X=\frac{1}{2}\left(1+(-1)^{|X|+R(L)}\right)$ where we define $R(L)=1$ when $L=0\mod{4}$, and $R(L)=0$ otherwise. 

Let us begin with computing $|K_{\emptyset}|$ for which all the factors of the form $(-1)^{x_j^{k'}}$ in \eqref{KXP} vanish. Observe that each cosine-factor in the product series over $j$ only can attain three different values, $\pm 1$ and $0$. Therefore, for given $\bar{x}$, we can only have a non-zero contribution to $K_{\emptyset}$ when all the cosine-factors in the product series take values $\pm 1$. Thus, there will be a contribution to $K_{\emptyset}$ only from half of the allowed strings of measurement settings per constituent Bell experiment. Let $C$ of the $2^{L-1}$ contributing strings (per Bell experiment) be the total number of strings that lead to a positive contribution to $K_{\emptyset}$. Then, the number of strings making a negative contribution is $2^{L-1}-C$. We can therefore write $|K_{\emptyset}|$ as  
\begin{equation}\label{Kform}
|K_{\emptyset}|=\frac{1}{2^{nL}}\times \left(2\max\{C,2^{L-1}-C\}-2^{L-1}\right)^n.
\end{equation}

Let us distinguish between when $R(L)=1$, in which case we label $C\rightarrow C_1$ and when $R(L)=0$ in which case we label $C\rightarrow C_0$. It is straightforward to compute $C_0$ and $C_1$. When $R(L)=1$, we have a positive contribution to $K_{\emptyset}$ from every string $x_j^1...x_j^L$ in which the entry $1$ appears a multiple of four number of times:   
\begin{multline}
C_1= \sum_{j=0}^{\lfloor L/4 \rfloor}\binom{L}{4j}=\\
2^{L/2-2}\left(2^{L/2}+\cos\left(\frac{\pi L}{4}\right)+(-1)^L\cos\left(\frac{3\pi L}{4}\right)\right).
\end{multline}
Similarly, when $R(L)=0$, we have a positive contribution to $K_{\emptyset}$ whenever the entry $1$ appears in $x_j^1...x_j^L$ a number of times which takes the form $4j+3$ for some non-negative integer $j$: 
\begin{multline}
C_0= \sum_{j=0}^{\lfloor (L-3)/4 \rfloor}\binom{L}{4j+3}=\\
2^{L/2-2}\left(2^{L/2}-\sin\left(\frac{\pi L}{4}\right)+(-1)^L\sin\left(\frac{3\pi L}{4}\right)\right).
\end{multline}
However, we can simplify the above by noting that if $L=2\mod{4}$, that is $L=4k+2$ for some non-negative integer $k$, then $C_0$ can be written  $C_0=16^k-4^k(-1)^k$. Similarly, using $L=0\mod{4}$, that is $L=4k$ for some positive integer $k$, we find  $C_1=4^{k-1}\left(4^k+2(-1)^k\right)$. Inserting either of these expressions into \eqref{Kform}, we find that $|K_{\emptyset}|=1/\sqrt{2^{nL}}$ which is thus true for any even $L$. Performing the analog calculation for odd values of $L$, one will find that $|K_{\emptyset}|=1/\sqrt{2^{n(L+1)}}$.

In order to determine the value of all other $|K_X|$ for $X\neq \emptyset$, it is sufficient to note that the symmetries of \eqref{KXP} together with the conventions introduced for $y_X$, will lead to $|K_X|=|K_{\emptyset}|$ for all $X$. Therefore, we end up with
\begin{eqnarray}\label{Qviolation}
\sum_{X \in \mathbb{P}(\mathbb{N}_{L})} |K_X|^{1/n} = 2^{L}\times |K_{\emptyset}|^{1/n}= \sqrt{2^{2\lfloor \frac{L}{2}\rfloor }}. 
\end{eqnarray} 

A peculiarity is that our choice of measurements has led to a weaker violation for odd values of $L$ (and in fact no violation for $L=1$). However, the violation for odd $L$ can be improved by a modification of the measurements of the non-Bob observers. For such purpose, we keep Bob's measurements as \eqref{BobM} but change the measurements of all remaining observers to $(X+Y)/\sqrt{2}$ (labeled $0$) and $(X-Y)/\sqrt{2}$ (labeled $1$). For example, if we consider quantum correlations in $\mathscr{N}_2^3$ we find the probability distribution associated to a constituent Bell experiment, i.e. the analog of \eqref{P} takes the form
\begin{multline}
P(a_j^1...a_j^3,b|x_j^1...x_j^3,y_X)=\\
\frac{1}{16}\left[1+\frac{1}{\sqrt{2}}(-1)^{b+\sum_{k=1,2,3}\left(a_j^k+y_Xx_j^k\right)}\right].
\end{multline}
Substituting this into \eqref{Pdist} and computing $|K_X|$ in analogy with the above, one will find the violation $\sum_{X}|K_X|^{1/2}=\sqrt{2^3}\nleq 1$ which improves the previous result for odd $L$. It is straightforward to generalize this to arbitrary $(n,L)$ which will lead to the violation $\sum |K_X|^{1/n}=\sqrt{2^L}$ holding true for arbitrary $L$. 

\subsection{Noise tolerance of quantum correlations}\label{3a}
Let us briefly study the possibility to violate the inequality \eqref{Inequality1} by the above procedure in the presence of environments with white noise. That is, with some probability $p$ a source will emit the GHZ state while with the probability $1-p$ the emitted state is a random noise signal modeled by the fully mixed state $\textbf{1}/2^{L+1}$. Thus, the state emitted by the $j$'th source is 
\begin{equation}
|\phi_j\rangle =p_j|GHZ\rangle\langle GHZ| +(1-p_j)\frac{\bf{1}}{2^{L+1}}.
\end{equation}
The total visibility of the system is the product over the visibilities of each source i.e. $V=p_1...p_n$. Let us find the critical value of $V$, by which we mean the largest number $V$ such that we can no longer violate the inequality \eqref{Inequality1} by the above method.

Since all the quantities $K_X$ are linear combinations of conditional probabilities, the $V$-dependent value of $K_X$ scales linearly with $V$: $K_X(V)=K_X\times V$. Thus, the critical value of $V$ is found from solving $2^L\times \left(V/\sqrt{2^{nL}}\right)^{1/n}=1$ which returns $V_{crit}=2^{-\frac{1}{2}nL}$. The noise tolerance of the quantum correlations thus scales exponentially with the number of observers and sources in the network.

This result can be compared to the critical visibility in the analog standard Bell scenario in which we remove the center node Bob and the $n$ sources and instead introduce only a single source emitting an $nL$-particle state shared between the remaining observers. For such a scenario with all observers performing one of two two-outcome measurements the most popular Bell inequality is due to Mermin \cite{M90}. Analyzing quantum violations of Mermin's inequality in the presence of imperfect visibilities yields a critical visibility identical to what we have obtained on our network $\mathscr{N}_n^L$. However, the two scenarios are far from equivalent despite their common critical visibilities. Our assumption \eqref{LocalRealism} is a significantly stronger constraint than the assumption of local causality used in standard Bell inequalities. Evidently, not only does this stronger constraints affect the strengh of classical correlations, but it also translates into a stronger constraint on the strength of quantum correlations on $\mathscr{N}_n^L$ rendering the critical visibility the same as in Bell experiments.

\subsection{Entanglement swapping in the center node}
A reasonable question to ask is if we can increase the violation of our inequality by considering Bob making a more general measurement. Such a measurement could be to jointly measure the $n$ qubits at Bob's disposal in a basis of entangled states. This would cause the global state of the system, which initially is a tensor over independent sources, to become entangled i.e. the entanglement in the network has been swapped. The basis of entangled states can be taken as a set of $2^n$ GHZ-like states, indexed by a bit-string $b^1...b^n$, obtained from 
\begin{equation}\label{basis}
|\xi_{b^1...b^n}\rangle = Z^{b^1}\otimes X^{b^2}\otimes...\otimes X^{b^n}|GHZ\rangle,
\end{equation}
where $Z=|0\rangle\langle 0|-|1\rangle\langle 1|$. 

Thus, such a complete entanglement swapping measurement would return one of $2^n$ possible outputs $b^1...b^n\in\{0,1\}^n$. Our inequality \eqref{Inequality1} is currently not of this form. However, a minor modification of \eqref{Inequality1} can take the entanglement swapping scenario into account: simply replace the conditioning of $K_X$ on $y_X$  in \eqref{KX} with a condition on some suitable bit $\hat{b}_X$ obtained from manipulations of the outcomes $b^1...b^n$.

Case studies have been performed for analyzing the strength of correlations arising from such entanglement swapping strategies. In particular, we have considered $\mathscr{N}_2^2$ for which we have modified the inequality \eqref{Inequality1} such that we let the outcome of Bob associated to $K_{\emptyset}$ and $K_{\{1,2\}}$ be $b^1$ and similarly $b^2$ for $K_{\{1\}}$ and $K_{\{2\}}$. Our method is analogous to that outlined in \cite{TSCA14}: we modify the problem so that it can be treated as a semi-definite program (SDP). To this end, we have to overcome the problem of the classical set of correlations being non-convex. We therefore restrict to the convex subset obtained from enforcing the restriction $K_{\emptyset}=\cdots=K_{\{1,2\}}$. Also, we fix the measurements of all observers except Bob to $(X+Y)/\sqrt{2}$ and $(X-Y)/\sqrt{2}$, which makes the optimization linear and thus suitable for techniques relying on SDPs. An SDP has been run optimizing the left-hand-side of our inequality over the measurement of Bob. Unsurprisingly such an optimization returns a measurement of Bob projecting the two qubits in a basis of Bell states of the type \eqref{basis}. However, the violation is the same as when using the separable measurement for Bob as in the above analysis, i.e. we find $\sum_X |K_X|^{1/2}=2$. 
In addition several variations of the above have been tried: the condition $K_{\emptyset}=\cdots=K_{\mathbb{N}_L}$ has been varied i.e. optimization has been performed along different convex subsets in combination with changing the fixed measurements of the non-Bob observers. Furthermore, numerical methods not based on SDPs have been used in brute-force optimizations in which we fix the measurement of Bob to a projection onto a basis of Bell states and optimize over the measurements of the non-Bob observers. Since this optimization problem is nonlinear it can easily return a local maxima. Therefore, the optimization was performed many times with different initial conditions. Nevertheless, no improvement over the quantum violations from the previous subsection have been found. Our results fall in line with the work of \cite{TSCA14} for the star-network (corresponding to $L=1$) for which no gain was found over separable measurements by introducing a complete joint measurement on $n$ qubits.

\subsection{Transforming the inequality for $\mathscr{N}_n^L$ into an inequality for a Bell experiment}

The quantities $K_X$ can be written on a more compact form by introducing correlators over the outcomes of all observers, $R^{j,k}$, and Bob in $\mathscr{N}_n^L$ defined as $\langle B_{y_X} R^{1,1}_{x_1^1}...R^{n,L}_{x_n^L}\rangle \equiv \sum_{\bar{a},b} (-1)^{b+\sum_{j,k} a_j^k}P(\bar{a},b|\bar{x},y_X)$. We can therefore re-write the definition \eqref{KX} compactly as $K_X=\frac{1}{2^{nL}}\sum_{\bar{x}}g(X)\langle B_{y_X} R^{1,1}_{x_1^1}...R^{n,L}_{x_n^L}\rangle$.

Now, notice that the measurement chosen for Bob in the above analysis, namely \eqref{BobM}, is a tensor product over the Hilbert spaces associated to respective qubit in his possession. Therefore, Bob's observable factors 
\begin{equation}
B_{y_X}\equiv M_{0|y_X}-M_{1|y_X}=\left(\Pi^{y_X}_0-\Pi^{y_X}_1\right)^{\otimes n}=\bigotimes_{j=1}^{n} B_{y_X}^j.
\end{equation}
This leads to a factorization $\langle B_{y_X} R^{1,1}_{x_1^1}...R^{n,L}_{x_n^L}\rangle=\langle B_{y_X} R^{1,1}_{x_1^1}...R^{1,L}_{x_1^L}\rangle ... \langle B_{y_X} R^{n,1}_{x_n^1}...R^{n,L}_{x_n^L}\rangle$. Thus, we re-write $K_X$ as
\begin{equation}
K_X=\frac{1}{2^{nL}}\prod_{j=1}^{n}\sum_{x_j^1...x_j^L}(-1)^{\sum_{k'\in X} x_j^{k'}}\langle B_{y_X} R^{j,1}_{x_j^1}...R^{j,L}_{x_j^L}\rangle.
\end{equation}
However, with the invoked symmetries of the correlators being invariant over the choice of $j$, our inequality \eqref{Inequality1} takes the form
\begin{multline}\label{MerminType}
\sum_{X\in \mathbb{P}(\mathbb{N}_L)} |K_X|^{1/n}=\\
\sum_{X\in \mathbb{P}(\mathbb{N}_L)}\left|\frac{1}{2^L}\sum_{x^1...x^L}(-1)^{\sum_{k'\in X} x^{k'}}\langle B_{y_X} R^{1}_{x^1}...R^{L}_{x^L}\rangle\right|\leq 1,
\end{multline}
which only considers the measurements and outcomes of a single Bell experiment (associated to a single multipartite source). Observe that for $L=1$ this is equivalent to the CHSH inequality \cite{CHSH69}. This can be seen from the fact that the domain of the sum reduces to  $\mathbb{P}(\mathbb{N}_L)=\{\emptyset,\{1\}\ \}$. Then,
\begin{multline}
\sum_{X=\emptyset,\{1\}}\left|\frac{1}{2}\sum_{x^1=0,1}(-1)^{\sum_{k'\in X} x^{k'}}\langle B_{y_X} R^{1}_{x^1}\rangle\right|=\\
\frac{1}{2}\left|\langle B_{y_\emptyset} R^1_0\rangle+\langle B_{y_\emptyset} R^1_1\rangle\right| +\frac{1}{2}\left|\langle B_{y_{\{1\}}} R^1_0\rangle-\langle B_{y_{\{1\}}} R^1_1\rangle\right|\\
 \leq 1.
\end{multline}
Identifying the labels $y_{\emptyset}$ and $y_{\{1\}}$ with '$0$' and '$1$' respectively, we see that we have derived the CHSH inequality.

By a similar procedure we can also derive Mermin's inequality for three observers ($L=2$) as a special case of \eqref{MerminType} by removing the absolute value with a plus sign for $X=\emptyset,\{1\},\{2\}$ and with a minus sign for $X=\{1,2\}$, and reduce Bob's four measurements to only two by setting  $y_{\{2\}}=y_{\{1\}}$ and $y_{\{1,2\}}=y_\emptyset$. We will then find
\begin{multline}
K_\emptyset^{1/n}+K_{\{1\}}^{1/n}+K_{\{2\}}^{1/n}-K_{\{1,2\}}^{1/n}=\frac{1}{2}\big( \langle B_{\{1\}} R^1_0 R^2_{0}\rangle\\
+\langle B_{\emptyset} R^1_0 R^2_{1}\rangle +\langle B_{\emptyset} R^1_1 R^2_{0}\rangle-\langle B_{\{1\}} R^1_1 R^2_{1}\rangle\big)\leq 1.
\end{multline}
Identifying the labels $y_{\emptyset}$ and $y_{\{1\}}$ with '$0$' and '$1$' returns Mermin's inequality.

\subsection{Exploration of quantum correlations}
In this section we will explore the set of quantum correlations on $\mathscr{N}_n^L$. In particular we will study the violation of our inequality as formualted in \eqref{MerminType} for a family of measurements in the XY-plane of the Bloch sphere in order to see how the violations behave for non-optimal measurements.  

We introduce variable measurements for all observers except Bob for whom we keep the measurement \eqref{BobM}. The remaining observers, $R^k$, can now chose between two measurements each parametrized as   
\begin{eqnarray}
R^{k}_{x^k}=\cos\left(\theta_{x^k}\right)X+\sin\left(\theta_{x^k}\right)Y,
\end{eqnarray}
for $\theta_{x^k}\in [0,\pi/2]$ for $x^k=0,1$ i.e. the measurements can be in any direction on the unit disk of the Bloch sphere in the plane spanned by $X$ and $Y$.

The correlators are computed from 
\begin{eqnarray}\nonumber
\langle B_0 R^{1}_{x^1}...R^{L}_{x^L}\rangle=\Tr\left(\rho \bigotimes_{k=1}^{L} \left(\cos\left(\theta_{x^k}\right)X+\sin\left(\theta_{x^k}\right)Y\right)\otimes X\right)\\\nonumber
\langle B_1 R^{1}_{x^1}...R^{L}_{x^L}\rangle=\Tr\left(\rho \bigotimes_{k=1}^{L} \left(\cos\left(\theta_{x^k}\right)X+\sin\left(\theta_{x^k}\right)Y\right)\otimes Y\right)\\
\end{eqnarray}
where we have taken $\rho$ to be the GHZ state.

The GHZ state is an eigenstate of strings of $L+1$ tensors of $X$ and $Y$ constrained such that $Y$ appears an even number of times in the string. Also, the sign of the associated eigenvalue $\pm 1$ is determined by whether $Y$ appears a number of times equal to zero or two modulo four. Using this property together with known trigonometric formulas, we find that
\begin{eqnarray}
\langle B_{y_X} R^{1}_{x^1}...R^{L}_{x^L}\rangle=\cos\left(\sum_{k=1}^{L}\theta_{x^k}+\frac{\pi y_X}{2}\right),
%\langle A_1 R^{1}_{x^1}...R^{L}_{x^L}\rangle=-\sin\left(\sum_{k=1}^{L}\theta_{x^k}\right),
\end{eqnarray}
where the convention for $y_X$ is the same as previously introduced.

Thus, all correlators associated to $k$ of the $L$ (non-Bob) observers making the measurement labeled $1$ are equal. We can compute the expression $|K_X|^{1/n}$, for any set $X$, by modifying a binomial expansion:
\begin{equation}\label{KXE}
K_X^{1/n}=\frac{1}{2^L}\sum_{k=0}^{L} \beta_k^{L,|X|}\cos\left(k\theta_1+(L-k)\theta_0+\frac{\pi y_X}{2}\right).
\end{equation}
However, the coefficients $\beta_k^{L,|X|}$ are only standard binomial coefficients in the case of $X=\emptyset$ since the pre-factors $(-1)^{x^k}$ in \eqref{MerminType} vanish. For arbitrary $X$, we denote the sum $s\equiv \sum_{r\in X}x^r$. For a particular value of $s$, we ask in how many ways one can arrange $k-s$ entries of $1$ among the remaining (non-fixed) $L-|X|$ measurement choices. Also, there are $\binom{|X|}{s}$ ways to chose a set $\{x^r\}_{r\in X}$ such that it sums to $s$. Thus, we are led to the expression
\begin{equation}\label{beta}
\beta_k^{L,|X|}=\sum_{s=0}^{\min\{|X|,k\}} (-1)^s \binom{L-|X|}{k-s}\binom{|X|}{s}.
\end{equation}
Since the values of $|K_X|^{1/n}$ now only depend on the cardinality $X$, we can write
\begin{equation}\label{KXEE}
\sum_{X\in \mathbb{P}(\mathbb{N}_L)} |K_X|^{1/n}=\sum_{|X|=0}^{L}\binom{L}{|X|}|K_X|^{1/n}\leq 1.
\end{equation}

Let us give an example. For simplicity, we introduce a convention $\theta\equiv \theta_0=\pi/2-\theta_1$ that allows us to easily recover the previously analyzed scenario in which $\theta_0=0$ and $\theta_1=\pi/2$. Considering the particular case of $L=2$, one can show that for even and odd values of $|X|$ one has
%\begin{gather}\nonumber\label{L=2}
%|K_{X_{|even}}|^{1/n}=\left|\frac{1}{2}\left(|X|-1\right)\left((|X|-1)\sin\left(2\theta\right)-1\right)\right|\\
%|K_{X_{|odd}}|^{1/n}=\left|\frac{1}{2}\left(|X|-2\right)|X|\cos\left(2\theta\right)\right|
%\end{gather} 
\begin{align}\nonumber\label{qqq}
|K_{\emptyset}|^{1/n}=\sin^2\left(\theta+\frac{\pi}{4}\right)\\\nonumber
|K_{\{1\}}|^{1/n}=|K_{\{2\}}|^{1/n}=\left|\sin\left(\theta+\frac{\pi}{4}\right)\sin\left(\theta-\frac{\pi}{4}\right)\right|\\
|K_{\{1,2\}}|^{1/n}=\sin^2\left(\theta-\frac{\pi}{4}\right).
\end{align}
Computing the Bell expression in \eqref{KXEE}:
\begin{multline}\label{example}
|K_{\emptyset}|^{1/n}+|K_{\{1\}}|^{1/n}+|K_{\{2\}}|^{1/n}+|K_{\{1,2\}}|^{1/n}=\\
1+2\left|\sin\left(\theta+\frac{\pi}{4}\right)\sin\left(\theta-\frac{\pi}{4}\right)\right|\geq 1 \hspace{3mm} \forall \theta\in [0,\frac{\pi}{2}],
\end{multline}
we find that a violation is obtained for all $\theta$ except when $\theta=\pi/4$.

More generally, we have been unable to analytically simplify the left-hand-side of the inequality into a compact expression in terms of $L, \theta$ using equations \eqref{KXE} and \eqref{beta}. However, numerical sampling over $\theta$ and for $L=1,...,50$ strongly indicates that
\begin{equation}
\sum_{X\in \mathbb{P}(\mathbb{N}_L)} |K_X|^{1/n}=
\left\{
     \begin{array}{lr}
       \sqrt{2^{2\lfloor \frac{L}{2}\rfloor }}\cos^L\left(\theta\right) \hspace{3 mm} 0\leq \theta\leq \frac{\pi}{4}\\ 
        \sqrt{2^{2\lfloor \frac{L}{2}\rfloor }}\sin^L\left(\theta\right)  \hspace{3 mm} \frac{\pi}{4}\leq \theta\leq \frac{\pi}{2}
     \end{array}
   \right.
\end{equation} 
which is likely to hold true for arbitrary $L$. Observe that the expression \eqref{example} easily can be brought to this form.

\begin{figure}
\centering
\includegraphics[width=\columnwidth]{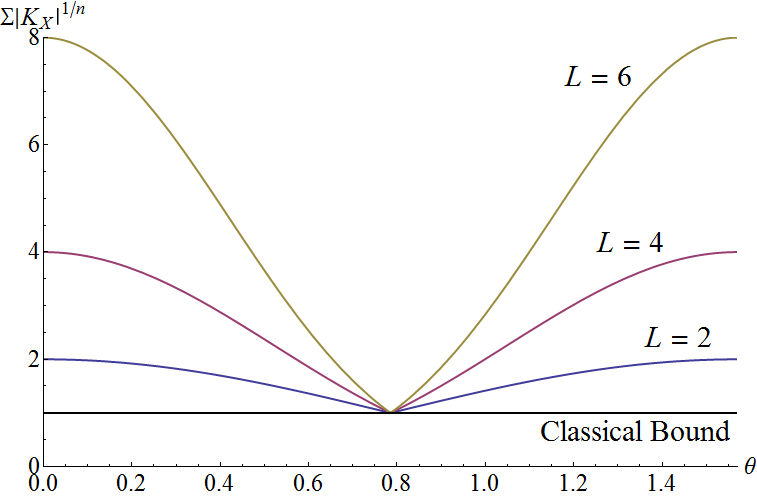}
\caption{The left-hand-side of inequality \eqref{KXEE} as a function of measurement parameter $\theta$ for $L=2,4,6$ obtained from quantum correlations versus the classical bound.}
\label{fig:1}
\end{figure}

\begin{figure}
\centering
\includegraphics[width=\columnwidth]{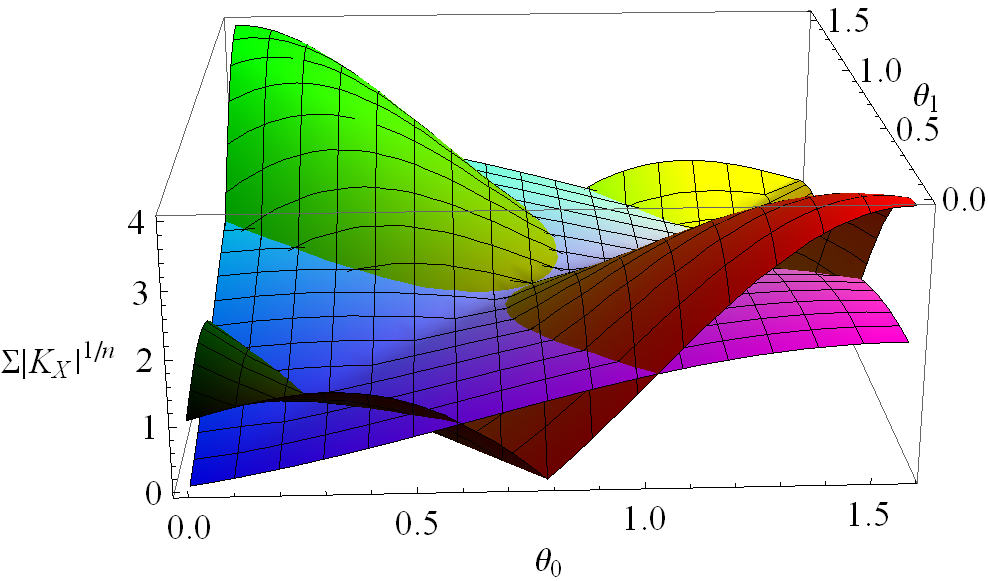}
\caption{The left-hand-side of inequality \eqref{KXEE} as a function of measurement parameters $\theta_0,\theta_1$ for $L=2,4$ obtained from quantum correlations.}
\label{fig:2}
\end{figure}

In figure \ref{fig:1} we have plotted the left-hand-side of the inequality \eqref{KXEE} as a function of $\theta$. 
Calculations similar to \eqref{qqq} but slightly more extensive have been used to extend the plot with $L=4,6$. We note that for the limited class of angles $\theta$ considered, we always have a violation of the inequality for the plotted cases except when $\theta=\pi/4$ which is expected since each observer always performs the same measurement. It appears very easy to find measurements violating the inequality. However, this is due to us making an optimal choice by setting $\theta_0=\pi/2-\theta_1$. This property disappears in we relax the constraint on $\theta_0,\theta_1$. For instance, in figure \ref{fig:2} we plot the left-hand-side of \eqref{KXEE} against both $\theta_0$ and $\theta_1$ for $L=2$ and $L=4$ from which it is clear that the relation between the measurement settings parametrized by $\theta_0,\theta_1$ and the violation of the inequality is not trivial. 

\section{Generalization to a broader class of networks}
\begin{figure}
\centering
\includegraphics[width=0.9\columnwidth]{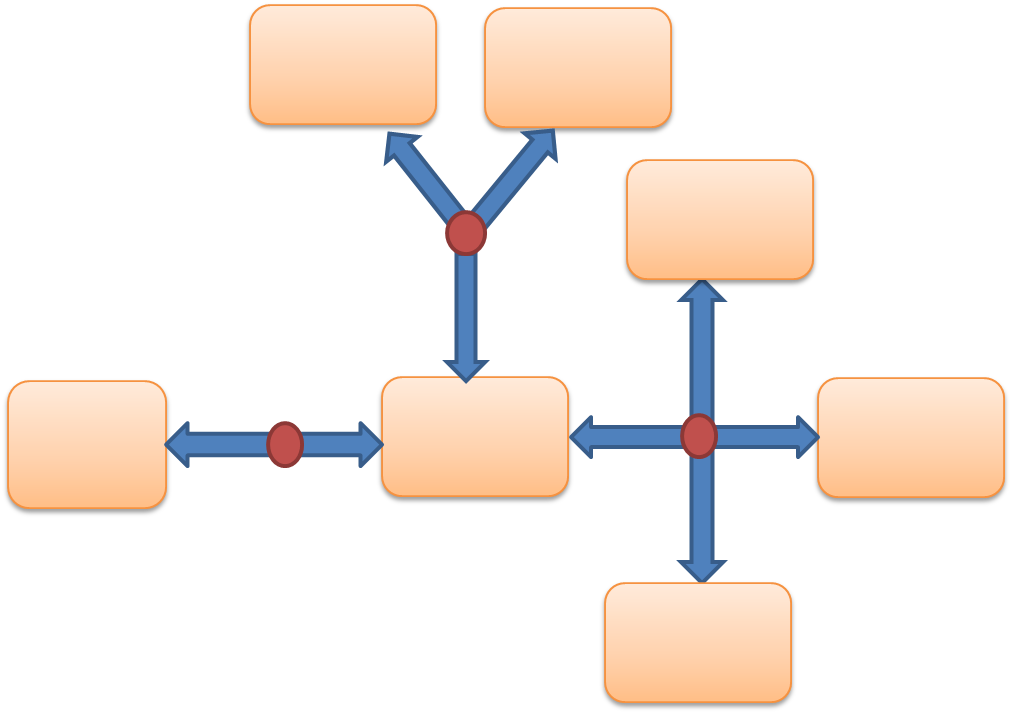}
\caption{The network $\mathscr{N}_3^{1,2,3}$.}
\label{FigMult}
\end{figure}

Let us now briefly consider a generalization of the analysis of correlations on $\mathscr{N}_n^L$ to include a broader class of networks. So far, we have let every source in the network emit a system of $L+1$ qubits so that one qubit is sent to Bob and each of the remaining $L$ qubits to a distinct observer. We will now relax this condition and let the $j$'th source emit an arbitrary number, $L_j+1$, of qubits such that one qubit is sent to Bob and the remaining $L_j$ sent to distinct observers. This new family of networks can be intuitively understood as connecting $n$ Bell experiments through a central node such that the $j$'th Bell experiment involves $L_j+1$ observers. We can charactherize the configuration of the network by the numbers $(n,L_1,...,L_n)$ and we abbreviate the network as $\mathscr{N}_n^{L_1\ldots L_n}$. In figure \ref{FigMult} we examplify this by illustrating the network $\mathscr{N}_3^{1,2,3}$.

To extend our inequalities \eqref{Inequality1} to account for correlations on $\mathscr{N}_n^{L_1\ldots L_n}$, we will define quantities analogous of those in \eqref{KX}: for every subset $X$ of $\mathbb{N}_L=\{1,...,L\}$ with $L\equiv \max\{L_1,...,L_n\}$ we define
\begin{equation}
Q_X=\frac{1}{2^{\sum_{j=1}^n L_j}}\sum_{\bar{x}} h(X)\langle B_{y_X} R^{1,1}_{x_1^1}\ldots R^{1,L_1}_{x_1^{L_1}}\ldots R^{n,L_n}_{x_n^{L_n}}\rangle
\end{equation}
where the $h(X)=\prod_{j=1}^{n}(-1)^{\sum_{k\in X_j}(-1)^{x_j^k}}$ and $X_j=\{s\in X| s\leq L_j\}$. Note that when $L_1=\ldots=L_n$, $Q_X$ reduces to $K_X$.

To bound $|Q_X|$, we can use a direct analogy of the technique used to bound the quantities $|K_X|$ in \eqref{bound}. Using \eqref{lemma}, we will then be led to the Bell-type inequality
\begin{equation}\label{ineq}
\sum_{X\in \mathbb{P}(\mathbb{N}_L)} |Q_X|^{1/n}\leq 2^L \times 2^{-\frac{1}{n}\sum_{j=1}^{n}L_j},
\end{equation} 
where the classical bound arises from the fact that for every $X_j$ there exists $2^{L-L_j}$ different sets $X\subset \mathbb{N}_L$ that contain $X_j$. Again, note that when $L_1=\ldots=L_n$, the classical bound reduces to that of inequality \eqref{Inequality1}, namely one.

The analysis of the quantum violations of inequality \eqref{ineq} is again a straightforward modification of our previous analysis in section \ref{sec3}. We will need to impose measurements of Bob that are somewhat different from \eqref{BobM}. Let us note that there can be at most $L$ different elements appearing in the set $\{L_1,...,L_n\}$, and let us denote these elements by $r_1,...,r_l$ for some $l\leq L$. Also, let $y_1\ldots y_l$ be an $l$-bit string. Bob's measurements are defined as
\begin{equation}\label{BobMM}
M_{b|y_1\ldots y_l}=\sum_{b_1\oplus \ldots \oplus b_n=b} \Pi^{y_1}_{b_1}\otimes \ldots \otimes \Pi^{y_1}_{b_{r_1}}\otimes \ldots \otimes \Pi^{y_L}_{b_{r_1+\ldots +r_l}}. 
\end{equation} 
In the case of $l=1$, we recover the two measurements \eqref{BobM}. More generally, there will be $2^l$ different measurements of Bob. 

Since the measurements \eqref{BobMM} are separable, the resulting joint probability distribution in $\mathscr{N}_n^{L_1\ldots L_n}$  will be subjected to a decomposition analogous to that in \eqref{Pdist} and thus the analysis reduces to considering the probability distribution associated to each source separately. If we distribute the state $\frac{1}{\sqrt{2}}(|0\rangle^{\otimes L_j+1}+|1\rangle^{\otimes L_j+1})$ in the $j$'th source, and let the non-Bob parties perform measurements $(X+Y)/\sqrt{2}$ or $(X-Y)/\sqrt{2}$, we will be lead to strong quantum violations analogous to those obtained in \ref{sec3}: $|Q_X|=2^{-\frac{1}{2}\sum_{j=1}^{n} L_j}$ leading to the violation of the inequality,
\begin{equation}
\sum_{X\in\mathbb{P}(\mathbb{N}_L)} |Q_X|^{1/n}=2^L\times 2^{-\frac{1}{2n}\sum_{j=1}^n L_j}
\end{equation}   
which is clearly a violation of \eqref{ineq} for all $\{L_1,...,L_n\}$.

Let us extend the analysis of section \ref{3a} by studying the quantum correlations when exposed to white noise. The $j$'th source emits a state 
\begin{equation}
|\phi_j\rangle =p_j|GHZ\rangle\langle GHZ| +(1-p_j)\frac{\bf{1}}{2^{L_j+1}} 
\end{equation}
for some $p_j\in [0,1]$ for $j=1,...,n$. The total visibility is defined as $V=p_1\ldots p_n$. 

Since $Q_X$ is a linear combination of conditional probabilities, we have that $Q_X(V)=V\times Q_X$. Therefore, the largest value of $V$ such that the above quantum correlations no longer violate the inequality \eqref{ineq} is found from solving $2^L\times (V\times 2^{-\frac{1}{2}\sum_{j=1}^{n}L_j})^{1/n}=2^L\times 2^{-\frac{1}{n}\sum_{j=1}^{n}L_j}$. The solution is  $V_{crit}=2^{-\frac{1}{2}\sum_{j=1}^{n}L_j}$. Thus, the results fall in line with those of section \ref{3a}: the white noise tolerance of the quantum correlations scales exponentially with the number of observers and sources in the network. This ought to confirm the intuition gained from considering $n$ Bell experiments connected through a separable action in the center node.

\section{conclusion}
We have studied correlations in a class of network configurations which can be understood as connecting many independent multipartite Bell experiments through a central node. We have derived tight Bell-type inequalities for such networks. By investigating quantum models of the arising probabilities, we found strong violations of our inequalities scaling exponentially with the number of observers connected to a source. However, the violation was found to be independent of the number of sources in the network. Somewhat surprisingly, we did not manage to increase the violation of our inequalities by performing an entanglement swapping measurement at the center node, which intuition suggests should offer a stronger result than performing a separable measurement. If entanglement swapping would have yielded an enhancement, then that would have constituted the sought for advantage over Bell experiments. We note that also previous works have studied network correlations in unsuccessful attempts to find an advantage over Bell experiments \cite{BRGP12, TSCA14}. In this sense, also our networks with multipartite sources failed in finding the sort of advantage anticipated from \cite{CASA10}: the quantum correlations found on the studied networks were not stronger (more resistant to noise) than the analog quantum correlations in Bell experiments. In the original work finding the advantage of networks \cite{CASA10}, a particular scenario is considered similar to the networks studied here in which the advantage over a Bell experiment on a single copy of a state appears in a network with seven sources in which a center node performing entanglement swapping. Therefore, it may be the case that an advantage over Bell experiments can be found for the networks analyzed here in the case of Bob performing an entanglement swapping measurement if there are sufficiently many sources in the network. However, due to limited available computation power, this could not be efficienctly investigated with numerics for networks of somewhat large size. 

Our negative result also suggests that it may be necessary to look for advantages over Bell experiments in other types of correlation experiments. It is interesting to study correlations in networks in which all observers perform measurements with more than two outcomes. For such purposes, novel techniques going beyond those presented here will most likely be necessary.

\section{Acknowledgements}
The author thanks Paul Skrzypczyk, Daniel Cavalcanti and Antonio Ac\'in for discussions.


\begin{thebibliography}{1}


\bibitem{Bell64}
	J. S. Bell, 
	\textit{On the Einstein-Podolsky-Rosen paradox},
	Physics (College. Park. Md). \textbf{1}, 195 (1964). 
	
\bibitem{Bell14}
	N. Brunner, D. Cavalcanti, S. Pironio, V. Scarani, and
	S. Wehner, 
	\textit{Bell nonlocality},
	Rev. Mod. Phys. 86, 419–478 (2014).

\bibitem{M90}
N. D. Mermin,
\textit{Extreme quantum entanglement in a superposition of macroscopically distinct states},
Phys. Rev. Lett. \textbf{65}, 1838 (1990).




\bibitem{BGP10}
	C. Branciard, N. Gisin and S. Pironio,
	\textit{Characterizing the nonlocal correlations created via entanglement swapping}, 
	Phys. Rev. Lett. \textbf{104}, 170401 (2010). 

\bibitem{BRGP12}
	C. Branciard, D. Rosset, N. Gisin and S. Pironio,
	\textit{Bilocal versus non-bilocal correlations in entanglement swapping experiments}, 
	Phys. Rev. A \textbf{85}, 032119 (2012).
	
\bibitem{MPS15}
K. Mukherjee, B. Paul, and D. Sarkar,
\textit{Correlations in n-local scenario},
arXiv: 1411.4188.

	
\bibitem{TSCA14}
	A. Tavakoli, P. Skrzypczyk, D. Cavalcanti and A. Ac\'in,
	\textit{Nonlocal correlations in the star-network configuration}
	Phys. Rev. A \textbf{90}, 062109 (2014).
	 
\bibitem{RB15}
D. Rosset, C. Branciard, T. J. Barnea, G. Pütz, N. Brunner and N. Gisin,
\textit{Nonlinear Bell inequalities tailored for quantum networks},
arXiv:1506.07380.

\bibitem{C15}
R. Chaves,
\textit{Polynomial Bell inequalities},
arXiv:1506.04325.

	
\bibitem{CAP10}
G. Chiribella, G. M. D’Ariano, and P. Perinotti,
\textit{Probabilistic theories with purification},
Phys. Rev. A \textbf{81}, 062348 (2010).



\bibitem{F12(1)}  
	T. Fritz,
	\textit{Beyond Bell's theorem: correlations scenarios},
	 New J. Phys. \textbf{14} 103001 (2012).	
	 
\bibitem{F12(2)}
	T. Fritz,
	\textit{Beyond Bell's Theorem II: Scenarios with arbitrary causal structure},	
	 arXiv: 1404.4812.

	
\bibitem{LS13}
M. S. Leifer and R. W. Spekkens,
\textit{Towards a formulation of quantum theory as a causally neutral theory of Bayesian inference},
Phys. Rev. A \textbf{88}, 052130 (2013).

\bibitem{CLM14}
R. Chaves, L. Luft, T. O. Maciel, D. Gross, D. Janzing, B. Schölkopf,
\textit{Inferring latent structures via information inequalities}
arXiv: 1407.2256.


\bibitem{HLP14}
J. Henson, R. Lal, and M. F. Pusey, 
\textit{Theory-independent limits on correlations from generalized Bayesian networks}, 
New J. Phys. 16, 113043 (2014).

\bibitem{LS15}
C. M. Lee and R. W. Spekkens,
\textit{Causal inference via algebraic geometry: necessary and sufficient conditions for the feasibility of discrete causal models},
arXiv:1506.03880.

\bibitem{CK15}
R. Chaves, R. Kueng, J. B. Brask, and D. Gross,
\textit{Unifying framework for relaxations of the causal assumptions in Bell’s theorem},
Phys. Rev. Lett. \textbf{114}, 140403 (2015).

\bibitem{WS15}
C. J. Wood and R. W. Spekkens,
\textit{The lesson of causal discovery algorithms for quantum correlations: causal explanations of Bell-inequality violations require fine-tuning},
New J. Phys. \textbf{17}, 073020 (2015).

\bibitem{RMG15}
R. Chaves, C. Majenz and D. Gross,
\textit{Information theoretic implications of quantum causal structures},
Nat. Comm \textbf{6}, 5766 (2015).


\bibitem{DI}
A. Ac\'in, N. Brunner, N. Gisin, S. Massar, S. Pironio, and
V. Scarani, 
\textit{Device-independent security of quantum cryptography against collective attacks},
Phys. Rev. Lett. 98, 230501 (2007).


\bibitem{Rand}
S. Pironio, A. Ac\'in, S. Massar, A. Boyer de la Giroday, D. N. Matsukevich, P. Maunz, S. Olmschenk, D. Hayes, L. Luo, T. A. Manning and C. Monroe,
\textit{Random numbers certified by Bell’s theorem},
Nature \textbf{464}, 1021 (2010).

\bibitem{ZZHE93}
M. Zukowski, A. Zeilinger, M. A. Horne and A. K. Ekert,
\textit{‘‘Event-ready-detectors’’ Bell experiment via entanglement swapping},
Phys. Rev. Lett. \textbf{71}, 4287 (1993).

\bibitem{ACL07}
A. Ac\'in, I. Cirac and M. Lewenstein,
\textit{Entanglement percolation in quantum networks},
Nature Physics \textbf{3}, 256 (2007).


\bibitem{SSRG11}
N. Sangouard, C. Simon, H. de Riedmatten, and N. Gisin,
\textit{Quantum repeaters based on atomic ensembles and linear optics},
Rev. Mod. Phys. \textbf{83}, 33 (2011).

\bibitem{SB05}
A. Sen(De), U. Sen, C. Brukner, V. Buzek and M. Zukowski,
\textit{Entanglement swapping of noisy states: A kind of superadditivity in nonclassicality},
Phys. Rev. A \textbf{72}, 042310 (2005).

\bibitem{CASA10}
D. Cavalcanti, M. L. Almeida, V. Scarani and A. Ac\'in,
\textit{Quantum networks reveal quantum nonlocality},
Nature Communications \textbf{2}, 184 (2010).

\bibitem{CHSH69}
J. F. Clauser, M. A. Horne, A. Shimony and R. A. Holt,
\textit{Proposed experiment to test local hidden-variable theories},
Phys. Rev. Lett. \textbf{23}, 880 (1969).







\end{thebibliography}
\end{document}